%% file: Template_Regular.tex
\documentclass{article}
\usepackage{spconf,amsmath, graphicx}
\usepackage{hyperref}
\usepackage{diagbox}
\usepackage{multicol}
\usepackage{multirow}
\usepackage[table ]{ xcolor}
\usepackage{pgfplots}
\pgfplotsset{compat=1.16}
\usepgfplotslibrary{groupplots}
\usepackage{bm}
\usepackage{makecell,rotating}
\usepackage{stfloats}
\usepackage{pdfpages}
\usepackage{subcaption}
\usetikzlibrary{patterns}
\usetikzlibrary{shapes.geometric}

\newcommand{\etal}{\emph{et al.}\ }

\title{How Far Are We from Robust Voice Conversion: A Survey}
\name{
    Tzu-hsien Huang$^*$,
    Jheng-hao Lin$^*$,
    Chien-yu Huang,
    Hung-yi Lee
    \thanks{$^*$ These authors contributed equally.}
}
\address{College of Electrical Engineering and Computer Science, National Taiwan University, Taiwan
}
\begin{document}
\maketitle
\input{0_Abstract}

\input{1_Introduction}

\label{sec:intro}

\input{2_Voice_Conversion_Models}

\input{3_Experimental_Setup}

\input{4_Robustness_of_Voice_Conversion_Model}

\input{5_Ablation_Study}

\input{6_Speaker_Embedding_}

\input{7_Conclusion}

\input{8_Acknowledgement}

\bibliographystyle{IEEEbib}
\bibliography{refs}

\end{document}

%% file: 0_Abstract.tex
\begin{abstract}
Voice conversion technologies have been greatly improved in recent years with the help of deep learning, but their capabilities of producing natural sounding utterances in different conditions remain unclear.
In this paper, we gave a thorough study of the robustness of known VC models.
We also modified these models, such as the replacement of speaker embeddings, to further improve their performances.
We found that the sampling rate and audio duration greatly influence voice conversion.
All the VC models suffer from unseen data, but AdaIN-VC is relatively more robust.
Also, the speaker embedding jointly trained is more suitable for voice conversion than those trained on speaker identification.

\end{abstract}


%% file: 1_Introduction.tex
\section{Introduction}

Voice conversion (VC) techniques aim to convert the speaker characteristic of an utterance into that of the target speaker while preserving linguistic content information.
In previous work, paired data of speakers were required to achieve VC.
Recently, several models were proposed to utilize non-parallel data~\cite{1810.12656, 8553236, serra2019blow, 2006.04154, 7820786}.
DGAN-VC~\cite{chou2018multi} learns disentangled content and speaker information by adversarial training.
StarGAN-VC~\cite{kameoka2018stargan} adopts conditional input to achieve many-to-many speaker voice conversion.
However, both are restricted to performing VC among seen speakers during training.
Zero-shot approaches~\cite{Liu2018, inproceedings_2, unknown, inproceedings, 2004.11284} are then considered, where the models can perform VC among any speakers without fine-tuning.
AdaIN-VC~\cite{chou2019one} applies instance normalization to meet this requirement.
\textsc{AutoVC}~\cite{qian2019autovc} employs pretrained d-vector~\cite{6854363} and information bottleneck for this purpose.

Although VC techniques were getting more powerful recently, most of the papers train and evaluate their VC models on the same corpora (usually the VCTK Corpus~\cite{veaux2016superseded}) with similar recording conditions. 
However, in real applications, the recording conditions of the utterances can be very different from the training data.
Are the VC models nowadays robust enough for real-world applications?
The answer is probably no.
Huang \etal \cite{2005.08781} successfully performed adversarial attack on VC models, and this suggests the VC models nowadays may still be not robust enough.
However, we do not know how non-robust they are 
and what kinds of mismatch have impacts on them?

This is probably the first paper studying the robustness of VC models. 
We survey the robustness of three popular VC models with five frequently-used datasets in the following three aspects.
    \begin{itemize}
    \setlength\itemsep{.2em}
        \item How models perform with audios with different sampling rate, duration, or even unseen language.
        
        \item To figure out which modules of these models are crucial to voice conversion through ablation study.
        
        \item To examine the influence of speaker embeddings so as to figure out which embedding is suitable for VC.
    \end{itemize}

%% file: 2_Voice_Conversion_Models.tex
\section{Voice Conversion}
Here we first introduce VC models and speaker embeddings involved in this paper.
We surveyed models achieving disentangled speaker timbre and content information: DGAN-VC~\cite{chou2018multi}, AdaIN-VC~\cite{chou2019one}, and \textsc{AutoVC}~\cite{qian2019autovc}.
All these models are publicly available and thus can be reproduced easily.
For speaker embeddings, we surveyed i-vector~\cite{1420369}, d-vector~\cite{6854363}, and x-vector~\cite{8461375}, all of which perform well in speaker verification, and further introduced a new embedding, v-vector.

\begin{figure}[hbt!]
    \centering
    \includegraphics[width=\linewidth]{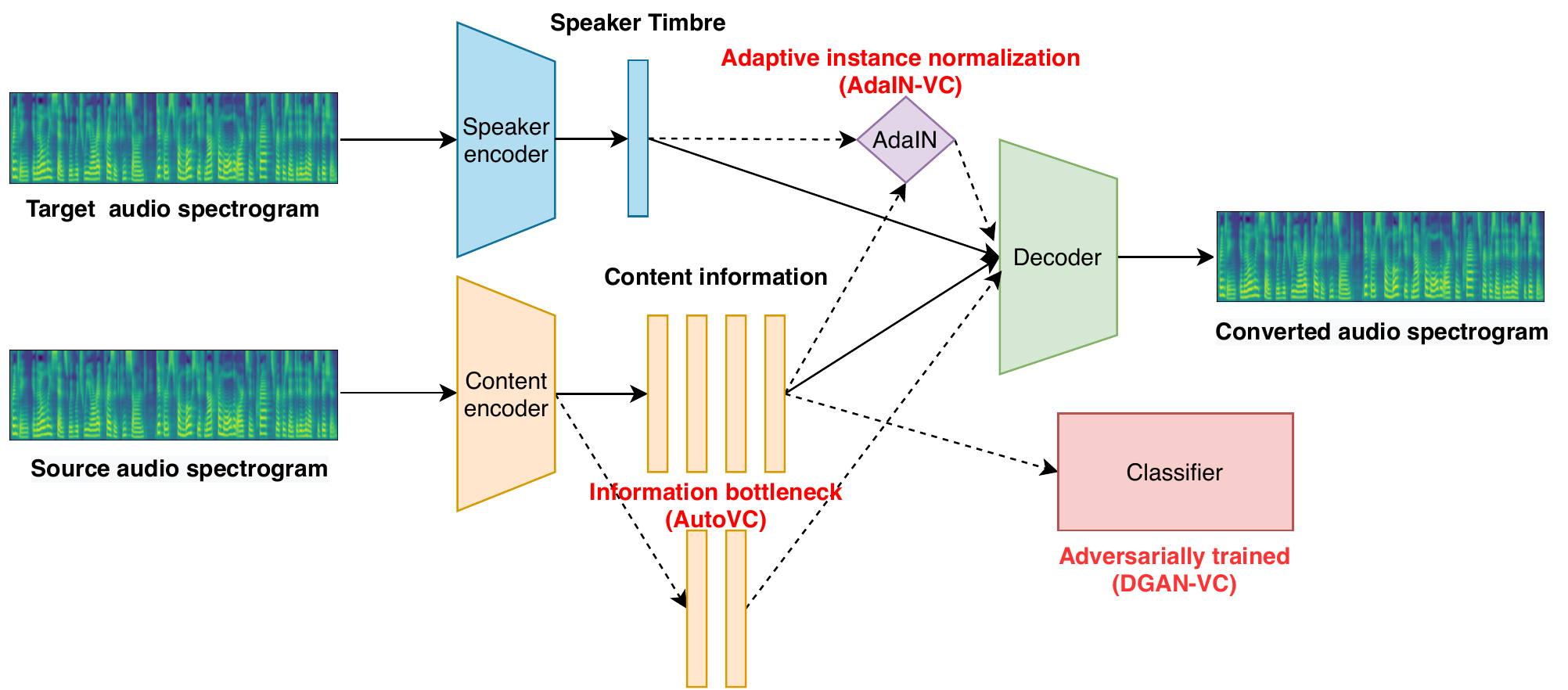}
    \caption{General framework of voice conversion models.}
    \label{fig:Architecture}
\end{figure}

\subsection{Voice conversion models}
    Recent VC models follow a general encoder-decoder architecture shown in Fig.~\ref{fig:Architecture}, in which timbre and linguistic content are extracted by speaker encoder ($E_{s}$) and content encoder ($E_c$) respectively.
    Decoder ($D$) then construct a converted signals conditioned on representations given by $E_s$ and $E_c$.
    
    DGAN-VC~\cite{chou2018multi} adopts a two-stage training framework for many-to-many voice conversion. 
    $E_{s}$ provides one-hot vectors as speaker embeddings to represent different speakers.
    In the first stage, an additional classifier is applied, taking the content representation as input and trained adversarially so as to obtain speaker-independent linguistic information.
    In the second stage, a generative adversarial network (GAN)~\cite{goodfellow2014generative} is then adopted to make the output spectra more realistic.
    
    AdaIN-VC~\cite{chou2019one} uses a variational autoencoder~\cite{kingma2013auto} (VAE), where latent representation is restricted by KL divergence loss, for content encoder and employs adaptive instance normalization~\cite{huang2017arbitrary} (AdaIN) to achieve zero-shot VC.
    Instance normalization used in $E_{c}$ removes timbre information while preserving the content information, and timbre information is then given by $E_{s}$ to the decoder with AdaIN layer.
    
    \textsc{AutoVC}~\cite{qian2019autovc} is an autoencoder with a carefully-designed information bottleneck trained with self-reconstruction loss only.
    The information bottleneck is the dimension of the latent vector between $E_{c}$ and $D$.
    With the bottleneck, linguistic content and timbre information are disentangled without explicit constraints, and the latter is provided by a pretrained d-vector during decoding.
    
\subsection{Speaker embeddings}
    Speaker embeddings, or speaker representations, are widely used in speaker verification and voice conversion, four types of which are involved in this paper.

    In i-vector~\cite{1420369}, the speaker and the utterance-dependent GMM supervector $\bm{M}$ are modeled as
    \begin{equation}
        \bm{M} = \bm{m} + \bm{T}\bm{w}
    \end{equation}
    where $\bm{m}$ denotes the speaker-independent GMM supervector and $\bm{T}$ is a low-rank total variability matrix which captures both speaker and utterance variability.
    And i-vector is simply the Maximum a Posterior point estimate of $\bm{w}$.
    
    D-vector~\cite{6854363} applies a DNN as speaker feature extractor and is trained on speaker identification task or with GE2E loss~\cite{wan2018generalized}.
    The last hidden layer's output is taken as a compact representation of a speaker, referred to as d-vector.
    
    X-vector~\cite{8461375} uses a time-delayed neural network (TDNN) to learn temporal context information.
    It is trained on speaker identification task, and the output of the last hidden layer is taken as the representation of a speaker, called x-vector.
    
    Additionally, v-vector is a new concept introduced in this paper, which is defined as the representation from speaker encoder jointly trained with VC models.

%% file: 3_Experimental_Setup.tex
\section{Experimental settings}

\subsection{Datasets}

We used CSTR VCTK Corpus~\cite{veaux2016superseded}, LibriTTS Corpus~\cite{zen2019libritts}, LibriSpeech ASR Corpus~\cite{panayotov2015librispeech} (LibriSpeech), CMU-ARCTIC databases~\cite{kominek2004cmu} (CMU) and THCHS-30~\cite{wang2015thchs} to evaluate performance of VC models.

For CSTR VCTK Corpus, we split it into two datasets called \textbf{VCTK-train} (99 speakers) and \textbf{VCTK-test} (10 speakers).
To avoid gender bias, we selected 5 male and 5 female speakers randomly as VCTK-test.
For LibriTTS Corpus, we used test-clean subset (\textbf{LibriTTS}).
For LibriSpeech ASR Corpus, we selected test-clean subset (\textbf{LibriSpeech}).
Table~\ref{tab:dataset_overview} lists detailed information of each dataset.
The abbreviations are used for simplicity in figures in later sections.

\begin{table}[htp]
    \centering
    \caption{An overview of the datasets.}
    \resizebox{\linewidth}{!}{
        \begin{tabular}{|c|c|c|c|c|}
            \hline
            \textbf{Dataset} & \textbf{Abbreviation} & \textbf{Speakers} & \textbf{Language} & \textbf{Sample rate} \\
            \hline
            VCTK-train & S & 99 & En & 48k \\
            \hline
            VCTK-test & U & 10 & En & 48k \\
            \hline
            LibriTTS & LT & 39 & En & 24k \\
            \hline
            LibriSpeech & LS & 40 & En & 16k \\
            \hline
            CMU & C & 18 & En & 16k\\
            \hline
            THCHS-30 & T & 60 & Zh & 16k \\
            \hline
        \end{tabular}
    }
    \label{tab:dataset_overview}
\end{table}

\subsection{Objective Metrics}
We consider two objective metrics: character error rate (CER) and speaker verification accept rate (SVAR).
The SVAR is the ratio of the number of utterances accepted by a speaker verification system to the total number of utterances.
A lower CER signifies a better content preserved, while a higher SVAR represents a more successful conversion of timbre.

Automatic speech recognition (ASR) measures how well linguistic information from the source utterance is preserved in the converted one.
We used Speech-to-Text service in google cloud speech API to compute CER.
As for Chinese, we decomposed the Chinese words into pinyin (spelled sounds) to avoid homophone problem, where several characters may pronounce the same but differ in meaning.

Speaker verification (SV) measures whether the converted utterance belongs to the speaker providing timbre information in VC.
We used a third-party pretrained speaker encoder\footnote{https://github.com/resemble-ai/Resemblyzer} to extract speaker embeddings from utterances.
A converted utterance is considered successfully converted if the cosine similarity of embeddings between the converted utterance and the target utterance (which provides timbre information in VC) is greater than a given threshold.
We obtained the threshold by computing equal error rate (EER) on datasets we used to test.
For each speaker, we randomly sampled 128 positive samples and 128 negative samples, and the total number of examples was more than 100k.
The EER is 5.65\% and the threshold is 0.6597.

\subsection{Implementation details}

\textsc{AutoVC}\footnote{https://github.com/auspicious3000/autovc}, AdaIN-VC\footnote{https://github.com/jjery2243542/adaptive\_voice\_conversion} and  DGAN-VC\footnote{https://github.com/jjery2243542/voice\_conversion} used here were obtained from official implementation and were all trained on VCTK-train set.
In their original implementations, DGAN-VC and AdaIN-VC utilized Griffin-Lim algorithm~\cite{griffin1984signal} to generate waveform from spectrogram while \textsc{AutoVC} applied WaveNet~\cite{van_den_Oord+2016} as vocoder.
To eliminate the influence of different vocoders, we modified each model to output 80-dim mel-spectrograms and adopted a pretrained MelGAN~\cite{kumar2019melgan} as vocoder to convert those mel-spectrograms into waveforms.
As DGAN-VC was limited to perform VC only among speakers seen during training in its original implementation, we replaced speaker embeddings of DGAN-VC with the speaker encoder architecture utilized in AdaIN-VC, which can encode utterance from unseen speaker, to meet zero-shot setting.

For speaker embeddings, i-vector and x-vector were obtained from the Kaldi~\cite{Povey:192584} pretrained systems trained on VoxCeleb1~\cite{Nagrani17} and Voxceleb2~\cite{Chung18b}.
As for d-vector, we made use of the pretrained d-vector model of \textsc{AutoVC} trained on VoxCeleb1 and LibriSpeech.
For v-vector, we use the pretrained speaker encoder from AdaIN-VC trained on VCTK-train.

%% file: 4_Robustness_of_Voice_Conversion_Model.tex
\section{Robustness of Voice Conversion Models}
In this section, we discuss the robustness of VC models in different scenarios. 
In Section~\ref{subsec:intra}, we investigated the situations that the training and testing utterances are from the same coups (VCTK) but with mismatches from different aspects including speakers, audio sample rates, and utterance durations.
In Section~\ref{subsec:noise}, we surveyed how models perform with noisy utterances.
In Section~\ref{subsec:inter}, we looked into how VC models perform when training and testing data are from different corpora with different characteristics (e.g., different languages).

\subsection{Intra-dataset} \label{subsec:intra}

\subsubsection{Experimental setup}

We created two new datasets from VCTK-train to model mismatch conditions. 
\textbf{VCTK-16} (S16) is downsampled from VCTK-train (from 48 kHz to 16 kHz); \textbf{VCTK-Long} (SL) is generated by concatenating several utterances of the same speaker in VCTK-train so that the duration is longer than 12 seconds.
The models in this subsection were trained on VCTK-train.
The models were evaluated under 8 scenarios (4 source sets $\times$ 2 target sets) listed in Table~\ref{tab:Intra-dataset-scenarios}.
In each scenario, we randomly sampled 1000 source-target utterance pairs from source set and target set to perform voice conversion.
As shown in Table~\ref{tab:Intra-dataset-scenarios}, X-Y denotes that the source utterance providing linguistic content is from corpus X, while the target utterance providing timbre information is from Y, or the VC models convert an utterance from X to make it sound like produced by the speaker in utterance in Y.
For S16-S and S16-U, we used the same pairs as which in S-S and S-U respectively.

\begin{table}[htb]
    \centering
    \caption{Scenarios of intra-dataset.}
    \resizebox{.75\linewidth}{!}{
        \begin{tabular}{|c|c|c|c|c|}
        \hline
        \backslashbox{\small Target}{\small Source} &  \textbf{S} & \textbf{U} & \textbf{S16} & \textbf{SL} \\
        \hline
        \textbf{S} & S-S & U-S & S16-S & SL-S \\
        \cline{1-5}
        \textbf{U} & S-U & U-U & S16-U & SL-U \\
        \hline
        \end{tabular}
    }
    \label{tab:Intra-dataset-scenarios}
\end{table}

\subsubsection{Results}
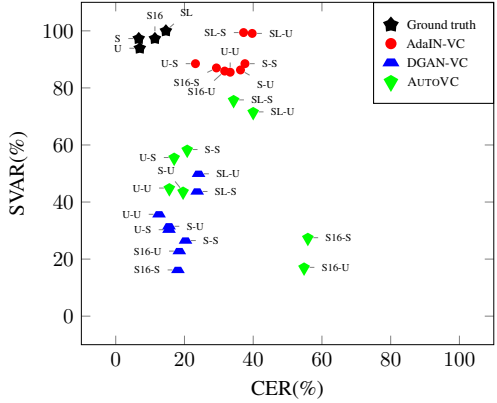
\begin{figure}[!t]
    \begin{tikzpicture}[pin distance=0.08cm, scale=0.8]
        \begin{axis}[
                xlabel={CER(\%)},
                ylabel={SVAR(\%)},
                xmin=-10, xmax=110,
                ymin=-10, ymax=110,
            ]
            \node[pin=180:{\tiny S}, star, black, fill, inner sep=1.5pt] at (axis cs:6.7, 97.3) {};
            \node[pin=180:{\tiny U}, star, black, fill, inner sep=1.5pt] at (axis cs:7.0, 93.9) {};
            \node[pin=90:{\tiny S16}, star, black, fill, inner sep=1.5pt] at (axis cs:11.4, 97.3) {};
            \node[pin=45:{\tiny SL}, star, black, fill, inner sep=1.5pt] at (axis cs:14.6, 100) {};
            
            \node[ pin=0:{\tiny S-S}, circle, red, fill, inner sep=1.5pt] at (axis cs:37.6, 88.5) {};
            \node[pin=180:{\tiny U-S}, circle, red, fill, inner sep=1.5pt] at (axis cs:23.2, 88.5) {};
            \node[pin=200:{\tiny S16-S}, circle, red, fill, inner sep=1.5pt] at (axis cs:29.3, 87.0) {};
            \node[pin=180:{\tiny SL-S}, circle, red, fill, inner sep=1.5pt] at (axis cs:37.2, 99.4) {};
            \node[pin=340:{\tiny S-U}, circle, red, fill, inner sep=1.5pt] at (axis cs:36.3, 86.3) {};
            \node[pin=90:{\tiny U-U}, circle, red, fill, inner sep=1.5pt] at (axis cs:33.3, 85.5) {};
            \node[pin=260:{\tiny S16-U}, circle, red, fill, inner sep=1.5pt] at (axis cs:31.7, 85.9) {};
            \node[pin=0:{\tiny SL-U}, circle, red, fill, inner sep=1.5pt] at (axis cs:39.7, 99.1) {};
            \node[pin=0:{\tiny S-S}, trapezium, blue, fill, inner sep=1.5pt] at (axis cs:20.3, 26.5) {};
            \node[pin=180:{\tiny U-S}, trapezium, blue,fill, inner sep=1.5pt] at (axis cs:15.5, 30.3) {};
            \node[pin=180:{\tiny S16-S}, trapezium, blue, fill, inner sep=1.5pt] at (axis cs:18.1, 16.2) {};
            \node[pin=0:{\tiny SL-S}, trapezium, blue, fill, inner sep=1.5pt] at (axis cs:23.7, 43.7) {};
            \node[pin=0:{\tiny S-U}, trapezium, blue, fill, inner sep=1.5pt] at (axis cs:15.5, 31.5) {};
            \node[pin=180:{\tiny U-U}, trapezium, blue, fill, inner sep=1.5pt] at (axis cs:12.6, 35.7) {};
            \node[pin=180:{\tiny S16-U}, trapezium, blue, fill, inner sep=1.5pt] at (axis cs:18.5, 22.8) {};
            \node[pin=0:{\tiny SL-U}, trapezium, blue, fill, inner sep=1.5pt] at (axis cs:24.1, 49.9) {};
            
            \node[pin=0:{\tiny S-S}, kite, green, fill, inner sep=1.5pt] at (axis cs: 20.8, 58.4) {};
            \node[pin=180:{\tiny U-S}, kite, green, fill, inner sep=1.5pt] at (axis cs:17.0, 55.6) {};
            \node[pin=0:{\tiny S16-S}, kite, green, fill, inner sep=1.5pt] at (axis cs:55.9, 27.5) {};
            \node[pin=0:{\tiny SL-S}, kite, green, fill, inner sep=1.5pt] at (axis cs:34.3, 75.8) {};
            \node[pin=95:{\tiny S-U}, kite, green, fill, inner sep=1.5pt] at (axis cs:19.6, 43.7) {};
            \node[pin=180:{\tiny U-U}, kite, green, fill, inner sep=1.5pt] at (axis cs:15.6, 44.9) {};
            \node[pin=0:{\tiny S16-U}, kite, green, fill, inner sep=1.5pt] at (axis cs:54.8, 17.1) {};
            \node[pin=0:{\tiny SL-U}, kite, green, fill, inner sep=1.5pt] at (axis cs:40.0, 71.6) {};
            
        \end{axis}
        \node [draw,fill=white] at (rel axis cs: 0.854, 0.852) {
            \begin{tikzpicture}
                \node[label=0:{\tiny Ground truth}, star, black, fill, inner sep=1.5pt] at (0, 0.75) {};
                \node[label=0:{\tiny AdaIN-VC}, circle, red, fill, inner sep=1.5pt] at (0, 0.5) {};
                \node[label=0:{\tiny DGAN-VC}, trapezium, blue, fill, inner sep=1.5pt] at (0, 0.25) {};
                \node[label=0:{\tiny \textsc{AutoVC}}, green, kite, fill, inner sep=1.5pt] at (0, 0) {}; 
            \end{tikzpicture}
        };
    \end{tikzpicture}
    \caption{Conversion results of intra-dataset. $S16$ and $SL$ represent VCTK-16 and VCTK-Long respectively.}
    \label{fig:intra-result}
\end{figure}

The results are shown in Fig.~\ref{fig:intra-result}.
Each point in the figure represents the performance of a VC model under a specific scenario in Table~\ref{tab:Intra-dataset-scenarios}.
Closer to the upper left corner means the better performance, and the closer distribution of a model means it is more robust, that is, less fluctuation under different scenarios.
Considering the performance of AdaIN-VC and \textsc{AutoVC}, we see that AdaIN-VC is better at converting timbre (higher SVAR), whereas \textsc{AutoVC} has clearer content (lower CER).
DGAN-VC is worse than \textsc{AutoVC} because they have similar CER, but DGAN-VC has lower SVAR.

For the robustness of the models, we see the results of AdaIN-VC cluster together, which suggests it is robust to both mismatched sampling rate and utterance duration.
As for DGAN-VC and \textsc{AutoVC}, the figure shows a mismatched sample rate may be harmful to both ASR and SV results.
Longer duration is beneficial for the conversion of speaker timbre with increasing SVAR, but it also increases CER.
The influence of sample rate and utterance duration is especially remarkable for \textsc{AutoVC}.
The observations suggest that both the sample rate and audio duration affect voice conversion, which should be noted.

\subsection{Noise} \label{subsec:noise}

\subsubsection{Experimental setup}
We used the same pairs as which in S-S in Table~\ref{tab:Intra-dataset-scenarios} and added the noises both on the source and target audios. 
We applied three noises: pink noise~\cite{doi:10.1111/j.1749-6632.1987.tb48734.x}, brownian noise, and indoor noise.
The pink and brownian noises are both generated at the level of roughly -30 dB, while we recorded indoor noise in our laboratory.
Ground-truth audios added with noises were also evaluated on both SV and ASR to discover whether objective metrics are robust to these noises.

\subsubsection{Results} 
\begin{figure}[!t]
    \centering
    \begin{tikzpicture}[pin distance=0.08cm, scale=0.8]
        \begin{axis}[
                xlabel={CER(\%)},
                ylabel={SVAR(\%)},
                xmin=-10, xmax=110,
                ymin=-10, ymax=110,
            ]
            \node[pin=270:{\tiny C}, star, black, fill, inner sep=1.5pt] at (axis cs:6.7, 97.3) {};
            \node[pin=0:{\tiny P}, star, black, fill, inner sep=1.5pt] at (axis cs:8.52, 98.72) {};
            \node[pin=180:{\tiny B}, star, black, fill, inner sep=1.5pt] at (axis cs:5.48, 99.98) {};
            \node[pin=45:{\tiny I}, star, black, fill, inner sep=1.5pt] at (axis cs: 6.48, 99.98) {};
            
            \node[pin=75:{\tiny C}, circle, red, fill, inner sep=1.5pt] at (axis cs:37.6, 88.5) {};
            \node[ pin=75:{\tiny P}, circle, red, fill, inner sep=1.5pt] at (axis cs:50.0, 51.7) {};
            \node[pin=75:{\tiny B}, circle, red, fill, inner sep=1.5pt] at (axis cs:48.8, 60.0) {};
            \node[pin=75:{\tiny I}, circle, red, fill, inner sep=1.5pt] at (axis cs:43.9, 80.8) {};
    
            \node[pin=75:{\tiny C}, trapezium, blue, fill, inner sep=1.5pt] at (axis cs:20.3, 26.5) {};
            \node[pin=75:{\tiny P}, trapezium, blue,fill, inner sep=1.5pt] at (axis cs:59.5, 4.1) {};
            \node[pin=75:{\tiny B}, trapezium, blue, fill, inner sep=1.5pt] at (axis cs:54.6, 3.7) {};
            \node[pin=75:{\tiny I}, trapezium, blue, fill, inner sep=1.5pt] at (axis cs:48.7, 4.4) {};
            
            \node[pin=75:{\tiny C}, kite, green, fill, inner sep=1.5pt] at (axis cs: 20.8, 58.4) {};
            \node[pin=75:{\tiny P}, kite, green, fill, inner sep=1.5pt] at (axis cs:66.0, 7.5) {};
            \node[pin=75:{\tiny B}, kite, green, fill, inner sep=1.5pt] at (axis cs:35.8, 29.6) {};
            \node[pin=75:{\tiny I}, kite, green, fill, inner sep=1.5pt] at (axis cs:24.8, 48.8) {};
            
        \end{axis}
        \node [draw,fill=white] at (rel axis cs: 0.854, 0.852) {
            \begin{tikzpicture}
                \node[label=0:{\tiny Ground truth}, star, black, fill, inner sep=1.5pt] at (0, 0.75) {};
                \node[label=0:{\tiny AdaIN-VC}, circle, red, fill, inner sep=1.5pt] at (0, 0.5) {};
                \node[label=0:{\tiny DGAN-VC}, trapezium, blue, fill, inner sep=1.5pt] at (0, 0.25) {};
                \node[label=0:{\tiny \textsc{AutoVC}}, green, kite, fill, inner sep=1.5pt] at (0, 0) {}; 
            \end{tikzpicture}
        };
    \end{tikzpicture}
    \caption{Conversion results under different noise scenarios. $C$, $B$, $P$ and $I$ represent clean audio, brownian noise, pink nosie and indoor noise respectively.}
    \label{fig:nosie}
\end{figure}
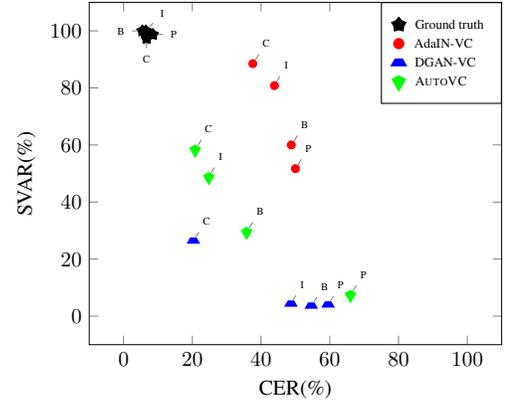

The results are shown in Fig.~\ref{fig:nosie}.
The Ground truth in the figure performed well on both CER and SVAR, whose results are at the upper-left corner of Fig.~\ref{fig:nosie}, suggesting that our objective metrics are resistant to these noises.
The performance of AdaIN-VC and \textsc{AutoVC} became worse with brownian noise and pink noise, and the degradation of \textsc{AutoVC} was more serious than that of AdaIN-VC.
In addition, pink noise is especially harmful to \textsc{AutoVC}.
The conversion of DGAN-VC failed under all the noisy inputs.
The results suggest AdaIN-VC is more robust to the noises than the others.

\subsection{Inter-dataset}  \label{subsec:inter}

\subsubsection{Experimental setup}
Here we utilized five datasets, including VCTK-train, LibriTTS, LibriSpeech, CMU, and THCHS-30.
The experiment was conducted in two cases: conversion from VCTK-train to the others and conversion from the others to VCTK-train.

\subsubsection{Results}
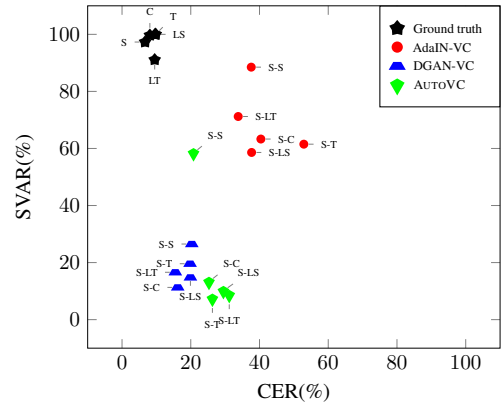
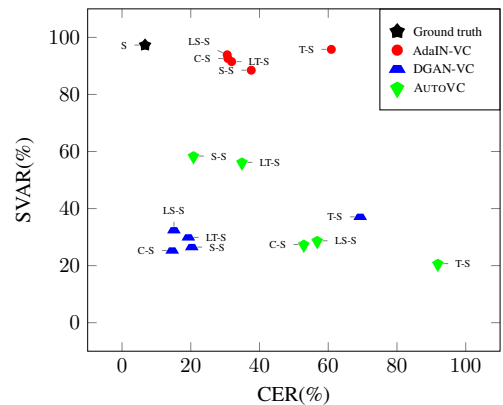
\begin{figure}[!h]
    \begin{subfigure}{.48\textwidth}
        \centering
        \begin{tikzpicture}[pin distance=0.08cm, scale=0.8]
            \begin{axis}[
                    xlabel={CER(\%)},
                    ylabel={SVAR(\%)},
                    xmin=-10, xmax=110,
                    ymin=-10, ymax=110,
                ]
                \node[pin=180:{\tiny S}, star, black, fill, inner sep=1.5pt] at (axis cs:6.7, 97.3) {};
                \node[pin=270:{\tiny LT}, star, black, fill, inner sep=1.5pt] at (axis cs:9.5, 91.1) {};
                \node[pin=0:{\tiny LS}, star, black, fill, inner sep=1.5pt] at (axis cs:9.64, 100) {};
                \node[pin=90:{\tiny C}, star, black, fill, inner sep=1.5pt] at (axis cs:8.1, 99.7) {};
                \node[pin=45:{\tiny T}, star, black, fill, inner sep=1.5pt] at (axis cs: 9.8, 100) {};
                
                \node[pin=0:{\tiny S-S}, circle, red, fill, inner sep=1.5pt] at (axis cs:37.6, 88.5) {};
                \node[pin=0:{\tiny S-LT}, circle, red, fill, inner sep=1.5pt] at (axis cs:33.8, 71.2) {};
                \node[ pin=0:{\tiny S-LS}, circle, red, fill, inner sep=1.5pt] at (axis cs:37.7, 58.6) {};
                \node[pin=0:{\tiny S-C}, circle, red, fill, inner sep=1.5pt] at (axis cs:40.4, 63.3) {};
                \node[pin=0:{\tiny S-T}, circle, red, fill, inner sep=1.5pt] at (axis cs:52.9, 61.5) {};
        
                \node[pin=180:{\tiny S-S}, trapezium, blue, fill, inner sep=1.5pt] at (axis cs:20.3, 26.5) {};
                \node[pin=180:{\tiny S-LT}, trapezium, blue, fill, inner sep=1.5pt] at (axis cs:15.5, 16.6) {};
                \node[pin=270:{\tiny S-LS}, trapezium, blue,fill, inner sep=1.5pt] at (axis cs:19.9, 14.8) {};
                \node[pin=180:{\tiny S-C}, trapezium, blue, fill, inner sep=1.5pt] at (axis cs:16.2, 11.3) {};
                \node[pin=180:{\tiny S-T}, trapezium, blue, fill, inner sep=1.5pt] at (axis cs:19.8, 19.6) {};
                
                \node[pin=45:{\tiny S-S}, kite, green, fill, inner sep=1.5pt] at (axis cs: 20.8, 58.4) {};
                \node[pin=270:{\tiny S-LT}, kite, green, fill, inner sep=1.5pt] at (axis cs: 31.2, 8.7) {};
                \node[pin=45:{\tiny S-LS}, kite, green, fill, inner sep=1.5pt] at (axis cs:29.5, 10.1) {};
                \node[pin=45:{\tiny S-C}, kite, green, fill, inner sep=1.5pt] at (axis cs:25.3, 13.3) {};
                \node[pin=270:{\tiny S-T}, kite, green, fill, inner sep=1.5pt] at (axis cs:26.3, 7.4) {};
                
            \end{axis}
            \node [draw,fill=white] at (rel axis cs: 0.854, 0.852) {
                \begin{tikzpicture}
                    \node[label=0:{\tiny Ground truth}, star, black, fill, inner sep=1.5pt] at (0, 0.75) {};
                    \node[label=0:{\tiny AdaIN-VC}, circle, red, fill, inner sep=1.5pt] at (0, 0.5) {};
                    \node[label=0:{\tiny DGAN-VC}, trapezium, blue, fill, inner sep=1.5pt] at (0, 0.25) {};
                    \node[label=0:{\tiny \textsc{AutoVC}}, green, kite, fill, inner sep=1.5pt] at (0, 0) {}; 
                \end{tikzpicture}
            };
        \end{tikzpicture}
        \vspace{-0.7em}
        \caption{VCTK-train to all datasets}
        \vspace{1em}
        \label{fig:VCTK-train to all datasets}
    \end{subfigure}
    \begin{subfigure}{.48\textwidth}
        \centering
        \begin{tikzpicture}[pin distance=0.08cm, scale=0.8]
            \begin{axis}[
                    xlabel={CER(\%)},
                    ylabel={SVAR(\%)},
                    xmin=-10, xmax=110,
                    ymin=-10, ymax=110,
            ]
                \node[pin=180:{\tiny S}, star, black, fill, inner sep=1.5pt] at (axis cs:6.7, 97.3) {};
                
                \node[pin=180:{\tiny S-S}, circle, red, fill, inner sep=1.5pt] at (axis cs:37.6, 88.5) {};
                \node[pin=0:{\tiny LT-S}, circle, red, fill, inner sep=1.5pt] at (axis cs:31.9, 91.5) {};
                \node[ pin=170:{\tiny LS-S}, circle, red, fill, inner sep=1.5pt] at (axis cs:30.6, 93.9) {};
                \node[pin=180:{\tiny C-S}, circle, red, fill, inner sep=1.5pt] at (axis cs:30.8, 92.6) {};
                \node[pin=180:{\tiny T-S}, circle, red, fill, inner sep=1.5pt] at (axis cs:60.9, 95.8) {};
            
                \node[pin=0:{\tiny S-S}, trapezium, blue, fill, inner sep=1.5pt] at (axis cs:20.3, 26.5) {};
                \node[pin=0:{\tiny LT-S}, trapezium, blue, fill, inner sep=1.5pt] at (axis cs:19.3, 29.9) {};
                \node[pin=90:{\tiny LS-S}, trapezium, blue,fill, inner sep=1.5pt] at (axis cs:15.1, 32.4) {};
                \node[pin=180:{\tiny C-S}, trapezium, blue, fill, inner sep=1.5pt] at (axis cs:14.6, 25.3) {};
                \node[pin=180:{\tiny T-S}, trapezium, blue, fill, inner sep=1.5pt] at (axis cs:69.3, 37.1) {};
                
                \node[pin=0:{\tiny S-S}, kite, green, fill, inner sep=1.5pt] at (axis cs: 20.8, 58.4) {};
                \node[pin=0:{\tiny LT-S}, kite, green, fill, inner sep=1.5pt] at (axis cs: 34.9, 56.2) {};
                \node[pin=0:{\tiny LS-S}, kite, green, fill, inner sep=1.5pt] at (axis cs:56.8, 28.7) {};
                \node[pin=180:{\tiny C-S}, kite, green, fill, inner sep=1.5pt] at (axis cs:52.9, 27.4) {};
                \node[pin=0:{\tiny T-S}, kite, green, fill, inner sep=1.5pt] at (axis cs:91.9, 20.7) {};
            
            \end{axis}
            \node [draw,fill=white,] at (rel axis cs: 0.854, 0.852) {
                \begin{tikzpicture}
                    \node[label=0:{\tiny Ground truth}, star, black, fill, inner sep=1.5pt] at (0, 0.75) {};
                    \node[label=0:{\tiny AdaIN-VC}, circle, red, fill, inner sep=1.5pt] at (0, 0.5) {};
                    \node[label=0:{\tiny DGAN-VC}, trapezium, blue, fill, inner sep=1.5pt] at (0, 0.25) {};
                    \node[label=0:{\tiny \textsc{AutoVC}}, green, kite, fill, inner sep=1.5pt] at (0, 0) {}; 
                \end{tikzpicture}
            };
        \end{tikzpicture}
        \vspace{-0.7em}
        \caption{All datasets to VCTK-train}
        \label{fig:All datasets to VCTK}
    \end{subfigure}
\caption{Conversion results of inter-dataset.}
\label{fig:results of inter-dataset}
\end{figure}

The results are shown in Fig.~\ref{fig:results of inter-dataset}.
There is little difference between the performance of DGAN-VC and \textsc{AutoVC} when conversions are performed between different datasets.
However, the performance of AdaIN-VC degrades slightly.
Fig.~\ref{fig:All datasets to VCTK} suggests that the performance of AdaIN-VC is stable with unseen speakers as source speaker, while the performance of \textsc{AutoVC} becomes worse drastically.

The results suggest that the carefully-designed information bottleneck in \textsc{AutoVC} disentangles content and speaker information successfully on the training set; however, it may not be well-generalized to unseen data.
The policy of tuning the size of the information bottleneck to make \textsc{AutoVC} robust is a critical question to be explored in the future.
The results of both from THCHS-30 to VCTK-train and from VCTK-train to THCHS-30 are quite different from others with higher CER and SVAR for all models.
This is due to the fact that THCHS-30 (Chinese) and VCTK-train (English) are in different languages, and the capability of cross-lingual VC is still limited for current VC models.

%% file: 5_Ablation_Study.tex
\section{Influence of model components}
In general, AdaIN-VC and \textsc{AutoVC} performed relatively better than DGAN-VC, and both have their own strengths.
We thus further study the two models in this section.

\begin{figure}[tb]
    \begin{subfigure}{.48\textwidth}
        \centering
        \begin{tikzpicture}[pin distance=0.08cm, scale=0.8]
            \begin{axis}[
                    xlabel={CER(\%)},
                    ylabel={SVAR(\%)},
                    xmin=-10, xmax=110,
                    ymin=-10, ymax=110,
                ]
                
                \node[pin=0:{\tiny S}, star, red, fill, inner sep=1.5pt] at (axis cs:20.8, 58.4) {};
                \node[pin=180:{\tiny U}, star, red, fill, inner sep=1.5pt] at (axis cs:15.6, 44.9) {};
                \node[pin=0:{\tiny LT}, star, red, fill, inner sep=1.5pt] at (axis cs:52.1, 9.6) {};
                \node[pin=0:{\tiny LS}, star, red, fill, inner sep=1.5pt] at (axis cs:72.9, 15.9) {};
                \node[pin=0:{\tiny C}, star, red, fill, inner sep=1.5pt] at (axis cs:56.0, 14.5) {};
                \node[pin=0:{\tiny T}, star, red, fill, inner sep=1.5pt] at (axis cs:95.3, 40.2) {};
        
                \node[pin=90:{\tiny S}, trapezium, blue, fill, inner sep=1.5pt] at (axis cs:62.9, 57.2) {};
                \node[pin=90:{\tiny U}, trapezium, blue, fill, inner sep=1.5pt] at (axis cs:60.7, 24.7) {};
                \node[pin=270:{\tiny LT}, trapezium, blue, fill, inner sep=1.5pt] at (axis cs:74.1, 1.6) {};
                \node[pin=90:{\tiny LS}, trapezium, blue,fill, inner sep=1.5pt] at (axis cs:77.9, 1.0) {};
                \node[pin=90:{\tiny C}, trapezium, blue, fill, inner sep=1.5pt] at (axis cs:72.8, 1.5) {};
                \node[pin=90:{\tiny T}, trapezium, blue, fill, inner sep=1.5pt] at (axis cs:100.0, 6.3) {};
                
                \node[pin=180:{\tiny S}, kite, pink, fill, inner sep=1.5pt] at (axis cs: 19.9, 21.5) {};
                \node[pin=180:{\tiny U}, kite, pink, fill, inner sep=1.5pt] at (axis cs: 16.8, 28.8) {};
                \node[pin=180:{\tiny LT}, kite, pink, fill, inner sep=1.5pt] at (axis cs: 43.6, 16.7) {};
                \node[pin=180:{\tiny LS}, kite, pink, fill, inner sep=1.5pt] at (axis cs: 51.0, 11.6) {};
                \node[pin=180:{\tiny C}, kite, pink, fill, inner sep=1.5pt] at (axis cs: 32.9, 9.1) {};
                \node[pin=180:{\tiny T}, kite, pink, fill, inner sep=1.5pt] at (axis cs: 88.1, 28.9) {};
                
                \node[pin=90:{\tiny S}, circle, green, fill, inner sep=1.5pt] at (axis cs: 29.3, 92.2) {};
                \node[pin=90:{\tiny U}, circle, green, fill, inner sep=1.5pt] at (axis cs: 16.3, 81.3) {};
                \node[pin=90:{\tiny LT}, circle, green, fill, inner sep=1.5pt] at (axis cs: 38.7, 62.6) {};
                \node[pin=90:{\tiny LS}, circle, green, fill, inner sep=1.5pt] at (axis cs: 76.2, 21.1) {};
                \node[pin=90:{\tiny C}, circle, green, fill, inner sep=1.5pt] at (axis cs: 67.5, 17.4) {};
                \node[pin=90:{\tiny T}, circle, green, fill, inner sep=1.5pt] at (axis cs: 96.1, 46.1) {};
                
            \end{axis}
            \node [draw,fill=white] at (rel axis cs: 0.825, 0.852) {
            \begin{tikzpicture}
                \node[label=0:{\tiny \textsc{AutoVC}}, star, red, fill, inner sep=1.5pt] at (0, 1) {};
                \node[label=0:{\tiny \textsc{AutoVC}-VAE}, trapezium, blue, fill, inner sep=1.5pt] at (0, 0.75) {};
                \node[label=0:{\tiny \textsc{AutoVC}-AdaIN}, kite, pink, fill, inner sep=1.5pt] at (0, 0.5) {};
                \node[label=0:{\tiny \textsc{AutoVC}-Vvec}, circle, green, fill, inner sep=1.5pt] at (0, 0.25) {}; 
            \end{tikzpicture}
            };
        \end{tikzpicture}
        \vspace{-0.7em}
        \caption{\textsc{AutoVC} modifications}
        \vspace{1em}
        \label{fig:AutoVC modifications}
    \end{subfigure}
    \begin{subfigure}{.48\textwidth}
        \centering
        \begin{tikzpicture}[pin distance=0.08cm, scale=0.8]
            \begin{axis}[
                    xlabel={CER(\%)},
                    ylabel={SVAR(\%)},
                    xmin=-10, xmax=110,
                    ymin=-10, ymax=110,
                ]
        
                \node[pin=0:{\tiny S}, star, red, fill, inner sep=1.5pt] at (axis cs: 37.6, 88.5) {};
                \node[pin=0:{\tiny U}, star, red, fill, inner sep=1.5pt] at (axis cs: 33.3, 85.5) {};
                \node[pin=90:{\tiny LT}, star, red, fill, inner sep=1.5pt] at (axis cs: 32.9, 86.5) {};
                \node[pin=180:{\tiny LS}, star, red, fill, inner sep=1.5pt] at (axis cs: 33.0, 85.3) {};
                \node[pin=270:{\tiny C}, star, red, fill, inner sep=1.5pt] at (axis cs: 29.9, 76.5) {};
                \node[pin=0:{\tiny T}, star, red, fill, inner sep=1.5pt] at (axis cs: 47.8, 98.4) {};
                
                \node[pin=0:{\tiny S}, circle, blue, fill, inner sep=1.5pt] at (axis cs: 24.5, 28.2) {};
                \node[pin=0:{\tiny U}, circle, blue, fill, inner sep=1.5pt] at (axis cs: 21.2, 38.1) {};
                \node[pin=0:{\tiny LT}, circle, blue, fill, inner sep=1.5pt] at (axis cs: 20.1, 46.9) {};
                \node[pin=0:{\tiny LS}, circle, blue, fill, inner sep=1.5pt] at (axis cs: 21.3, 51.3) {};
                \node[pin=0:{\tiny C}, circle, blue, fill, inner sep=1.5pt] at (axis cs: 18.8, 33.1) {};
                \node[pin=90:{\tiny T}, circle, blue, fill, inner sep=1.5pt] at (axis cs: 21.9, 92.1) {};
                
            \end{axis}
            \node [draw,fill=white] at (rel axis cs: 0.844, 0.907) {
                \begin{tikzpicture}
                    \node[label=0:{\tiny AdaIN-VC}, star, red, fill, inner sep=1.5pt] at (0, 0.5) {};
                    \node[label=0:{\tiny AdaIN-VC-AE}, circle, blue, fill, inner sep=1pt] at (0, 0.25) {}; 
                \end{tikzpicture}
            };
        \end{tikzpicture}
        \vspace{-0.7em}
        \caption{AdaIN-VC modifications}
        \label{fig:AdaIN-VC modifications}
    \end{subfigure}
\caption{Conversion results of various combinations of components.}
\label{fig:The influence of model components}
\end{figure}

\subsection{Experimental setup}
First, we trained \textsc{AutoVC} whose speaker encoder is replaced with v-vector from trained AdaIN-VC (\textbf{\textsc{AutoVC}-Vvec}).
Second, we added VAE architecture into \textsc{AutoVC} (\textbf{-VAE}).
Third, We integrated AdaIN layer into the decoder of \textsc{AutoVC} (\textbf{-AdaIN}).
Finally, we replaced the VAE architecture in AdaIN-VC with an autoencoder one (\textbf{AdaIN-VC-AE}).

\subsection{Results}
The results are shown in Fig.~\ref{fig:The influence of model components}.
\textsc{AutoVC}-Vvec performed better than \textsc{AutoVC} in terms of SVAR on all datasets.
However, \textsc{AutoVC}-Vvec did not show significant improvement in terms of CERs compared to \textsc{AutoVC}, so it is still not robust to unseen corpora.
The influence of speaker embeddings is discussed in Section~\ref{sec:speaker_embed}.

The performances of AdaIN-VC-AE and \textsc{AutoVC}-AdaIN were not ideal on both CER and SVAR, suggesting that using AdaIn layer alone is not powerful enough to convert speaker timbre.
\textsc{AutoVC}-VAE performed the worst in the aspect of CER among \textsc{AutoVC} modifications.
Both small size of latent vector and VAE architecture limit the information that the content encoder provides to the decoder.
The undesirable result of \textsc{AutoVC}-VAE might be attributed to insufficient content information induced by an improper combination of latent vector and VAE architecture.

\begin{figure}[t]
    \begin{subfigure}{.45\textwidth}
        \centering
        \begin{tikzpicture}[pin distance=0.08cm, scale=0.8]
            \begin{axis}[
                    xlabel={CER(\%)},
                    ylabel={SVAR(\%)},
                    xmin=-10, xmax=110,
                    ymin=-10, ymax=110,
                ]
                
                \node[pin=180:{\tiny S}, star, red, fill, inner sep=1.5pt] at (axis cs:20.8, 58.4) {};
                \node[pin=180:{\tiny U}, star, red, fill, inner sep=1.5pt] at (axis cs:15.6, 44.9) {};
                \node[pin=180:{\tiny LT}, star, red, fill, inner sep=1.5pt] at (axis cs:52.1, 9.6) {};
                \node[pin=225:{\tiny LS}, star, red, fill, inner sep=1.5pt] at (axis cs:72.9, 15.9) {};
                \node[pin=180:{\tiny C}, star, red, fill, inner sep=1.5pt] at (axis cs:56.0, 14.5) {};
                \node[pin=180:{\tiny T}, star, red, fill, inner sep=1.5pt] at (axis cs:95.3, 40.2) {};
        
                \node[pin=270:{\tiny S}, trapezium, blue, fill, inner sep=1.5pt] at (axis cs:47.0, 96.3) {};
                \node[pin=270:{\tiny U}, trapezium, blue, fill, inner sep=1.5pt] at (axis cs:42.3, 81.0) {};
                \node[pin=270:{\tiny LT}, trapezium, blue, fill, inner sep=1.5pt] at (axis cs:61.9, 51.3) {};
                \node[pin=270:{\tiny LS}, trapezium, blue,fill, inner sep=1.5pt] at (axis cs:67.3, 54.6) {};
                \node[pin=270:{\tiny C}, trapezium, blue, fill, inner sep=1.5pt] at (axis cs:59.5, 8.0) {};
                \node[pin=270:{\tiny T}, trapezium, blue, fill, inner sep=1.5pt] at (axis cs:100, 77.9) {};
                
                \node[pin=180:{\tiny S}, kite, pink, fill, inner sep=1.5pt] at (axis cs: 85, 62.9) {};
                \node[pin=90:{\tiny U}, kite, pink, fill, inner sep=1.5pt] at (axis cs: 84.4, 24.7) {};
                \node[pin=180:{\tiny LT}, kite, pink, fill, inner sep=1.5pt] at (axis cs: 82.9, 6.6) {};
                \node[pin=45:{\tiny LS}, kite, pink, fill, inner sep=1.5pt] at (axis cs: 88.6, 6.5) {};
                \node[pin=270:{\tiny C}, kite, pink, fill, inner sep=1.5pt] at (axis cs: 85.3, 1.3) {};
                \node[pin=45:{\tiny T}, kite, pink, fill, inner sep=1.5pt] at (axis cs: 100, 9.2) {};
                
                \node[pin=90:{\tiny S}, circle, green, fill, inner sep=1.5pt] at (axis cs: 29.3, 92.2) {};
                \node[pin=90:{\tiny U}, circle, green, fill, inner sep=1.5pt] at (axis cs: 16.3, 81.3) {};
                \node[pin=90:{\tiny LT}, circle, green, fill, inner sep=1.5pt] at (axis cs: 38.7, 62.6) {};
                \node[pin=90:{\tiny LS}, circle, green, fill, inner sep=1.5pt] at (axis cs: 76.2, 21.1) {};
                \node[pin=90:{\tiny C}, circle, green, fill, inner sep=1.5pt] at (axis cs: 67.5, 17.4) {};
                \node[pin=90:{\tiny T}, circle, green, fill, inner sep=1.5pt] at (axis cs: 96.1, 46.1) {};
                
                \node[pin=90:{\tiny S}, dart, yellow, fill, inner sep=1.5pt] at (axis cs: 61.8, 61.3) {};
                \node[pin=90:{\tiny U}, dart, yellow, fill, inner sep=1.5pt] at (axis cs: 61.9, 15.3) {};
                \node[pin=90:{\tiny LT}, dart, yellow, fill, inner sep=1.5pt] at (axis cs: 85.7, 6.5) {};
                \node[pin=270:{\tiny LS}, dart, yellow, fill, inner sep=1.5pt] at (axis cs: 96.4, 1.0) {};
                \node[pin=270:{\tiny C}, dart, yellow, fill, inner sep=1.5pt] at (axis cs: 89.3, 0.1) {};
                \node[pin=315:{\tiny T}, dart, yellow, fill, inner sep=1.5pt] at (axis cs: 98.9, 0.61) {};
                
            \end{axis}
            \node [draw,fill=white] at (rel axis cs: 0.172, 0.176) {
                \begin{tikzpicture}
                    \node[label=0:{\tiny \textsc{AutoVC}-Dvec}, star, red, fill, inner sep=1.5pt] at (0, 1) {};
                    \node[label=0:{\tiny \textsc{AutoVC}-Ivec}, trapezium, blue, fill, inner sep=1.5pt] at (0, 0.75) {};
                    \node[label=0:{\tiny \textsc{AutoVC}-Xvec}, kite, pink, fill, inner sep=1.5pt] at (0, 0.5) {};
                    \node[label=0:{\tiny \textsc{AutoVC}-Vvec}, circle, green, fill, inner sep=1.5pt] at (0, 0.25) {};
                    \node[label=0:{\tiny \textsc{AutoVC}-Joint}, dart, yellow, fill, inner sep=1.5pt] at (0, 0.) {}; 
                \end{tikzpicture}
            };
        \end{tikzpicture}
        \vspace{-0.7em}
        \caption{\textsc{AutoVC} modifications}
        \vspace{1em}
        \label{fig:AutoVC with different speaker embedding}
    \end{subfigure}
    \begin{subfigure}{.45\textwidth}
        \centering
        \begin{tikzpicture}[pin distance=0.08cm, scale=0.8]
            \begin{axis}[
                    xlabel={CER(\%)},
                    ylabel={SVAR(\%)},
                    xmin=-10, xmax=110,
                    ymin=-10, ymax=110,
                ]
                
                \node[pin=270:{\tiny S}, star, red, fill, inner sep=1.5pt] at (axis cs:24.6, 21.0) {};
                \node[pin=270:{\tiny U}, star, red, fill, inner sep=1.5pt] at (axis cs:17.7, 29.3) {};
                \node[pin=270:{\tiny LT}, star, red, fill, inner sep=1.5pt] at (axis cs:47.6, 29.3) {};
                \node[pin=270:{\tiny LS}, star, red, fill, inner sep=1.5pt] at (axis cs:57.1, 29.3) {};
                \node[pin=270:{\tiny C}, star, red, fill, inner sep=1.5pt] at (axis cs:37.5, 1.2) {};
                \node[pin=270:{\tiny T}, star, red, fill, inner sep=1.5pt] at (axis cs:91.0, 22.7) {};
        
                \node[pin=270:{\tiny S}, trapezium, blue, fill, inner sep=1.5pt] at (axis cs:31.6, 72.6) {};
                \node[pin=270:{\tiny U}, trapezium, blue, fill, inner sep=1.5pt] at (axis cs:24.4, 70.7) {};
                \node[pin=270:{\tiny LT}, trapezium, blue, fill, inner sep=1.5pt] at (axis cs:65.8, 30.3) {};
                \node[pin=270:{\tiny LS}, trapezium, blue,fill, inner sep=1.5pt] at (axis cs:60.8, 34.4) {};
                \node[pin=270:{\tiny C}, trapezium, blue, fill, inner sep=1.5pt] at (axis cs:31.0, 9.7) {};
                \node[pin=270:{\tiny T}, trapezium, blue, fill, inner sep=1.5pt] at (axis cs:78.9, 65.9) {};
                
                \node[pin=90:{\tiny S}, kite, pink, fill, inner sep=1.5pt] at (axis cs: 28.6, 42.2) {};
                \node[pin=90:{\tiny U}, kite, pink, fill, inner sep=1.5pt] at (axis cs: 18.3, 50.2) {};
                \node[pin=90:{\tiny LT}, kite, pink, fill, inner sep=1.5pt] at (axis cs: 23.3, 29.0) {};
                \node[pin=0:{\tiny LS}, kite, pink, fill, inner sep=1.5pt] at (axis cs: 27.8, 29.4) {};
                \node[pin=270:{\tiny C}, kite, pink, fill, inner sep=1.5pt] at (axis cs: 25.1, 3.7) {};
                \node[pin=90:{\tiny T}, kite, pink, fill, inner sep=1.5pt] at (axis cs: 53.5, 77.8) {};
                
                \node[pin=0:{\tiny S}, circle, green, fill, inner sep=1.5pt] at (axis cs: 37.6, 88.5) {};
                \node[pin=0:{\tiny U}, circle, green, fill, inner sep=1.5pt] at (axis cs: 33.3, 85.5) {};
                \node[pin=180:{\tiny LT}, circle, green, fill, inner sep=1.5pt] at (axis cs: 32.9, 86.5) {};
                \node[pin=225:{\tiny LS}, circle, green, fill, inner sep=1.5pt] at (axis cs: 33.0, 85.3) {};
                \node[pin=0:{\tiny C}, circle, green, fill, inner sep=1.5pt] at (axis cs: 29.9, 76.5) {};
                \node[pin=0:{\tiny T}, circle, green, fill, inner sep=1.5pt] at (axis cs: 47.8, 98.4) {};
                
            \end{axis}
            \node [draw,fill=white] at (rel axis cs: 0.8548, 0.852) {
                \begin{tikzpicture}
                    \node[label=0:{\tiny AdaIN-Dvec}, star, red, fill, inner sep=1.5pt] at (0, 1) {};
                    \node[label=0:{\tiny AdaIN-Ivec}, trapezium, blue, fill, inner sep=1.5pt] at (0, 0.75) {};
                    \node[label=0:{\tiny AdaIN-Xvec}, kite, pink, fill, inner sep=1.5pt] at (0, 0.5) {};
                    \node[label=0:{\tiny AdaIN-Joint}, circle, green, fill, inner sep=1.5pt] at (0, 0.25) {};
                \end{tikzpicture}
            };
        \end{tikzpicture}
        \vspace{-0.7em}
        \caption{AdaIN-VC modifications}
        \label{fig:AdaIN-VC with different speaker embedding}
    \end{subfigure}
\caption{Conversion results of different speaker embeddings.}
\label{fig:The Influence of Speaker Embeddings}
\end{figure}

%% file: 6_Speaker_Embedding_.tex
\section{Influence of Speaker Embeddings}\label{sec:speaker_embed}
In this section, we discuss how different speaker embeddings influence the performance of VC.

\begin{figure}[htb]
    \centering
    \resizebox{.95\linewidth}{!}{
    \begin{tikzpicture}
        \begin{groupplot}[
                group style={
                        group size=2 by 1,
                    },
            ]
            \nextgroupplot[
                    title={VCTK}, 
                    height=5cm, width=5cm,
                    ylabel={Similarity},
                    ylabel style={
                            yshift=-2mm,
                        },
                    ybar stacked,
                    ymin=0, ymax=100,
                    xtick=data,
                    symbolic x coords={AdaIN-VC, \textsc{AutoVC}, \textsc{AutoVC}-Vvec, Ground truth},
                    x tick label style={
                            rotate=45,
                            anchor=east,
                            font=\footnotesize
                        },
                    xtick style={
                            draw=none,
                        },
                    legend style={
                            at={(-0.1, -0.18)},
                            legend columns=2,
                            fill=none,
                            draw=black,
                            anchor=north,
                            legend cell              align=left,
                        },
                    legend to name=fred,
                ]
            \addplot[
                    ybar,
                    fill=blue,
                ] coordinates {
                    (AdaIN-VC, 22.45762712)
                    (\textsc{AutoVC}, 17.84232365)
                    (\textsc{AutoVC}-Vvec, 40.90909091)
                    (Ground truth, 68.3982684)
                };
            \addplot[
                    ybar,
                    fill=teal,
                ] coordinates {
                    (AdaIN-VC, 40.25423729)
                    (\textsc{AutoVC}, 21.57676349)
                    (\textsc{AutoVC}-Vvec, 25.20661157)
                    (Ground truth, 18.18181818)
                };
            \addplot[
                    ybar,
                    fill=orange,
                ] coordinates {
                    (AdaIN-VC, 19.91525424)
                    (\textsc{AutoVC}, 32.78008299)
                    (\textsc{AutoVC}-Vvec, 20.24793388)
                    (Ground truth, 09.956709957)
                };
            \addplot[
                    ybar,
                    fill=red,
                ] coordinates {
                    (AdaIN-VC, 17.37288136)
                    (\textsc{AutoVC}, 27.80082988)
                    (\textsc{AutoVC}-Vvec, 13.63636364)
                    (Ground truth, 03.463203463)
                };
            \addlegendentry{same, absolutely sure};    
            \addlegendentry{same, not sure};    
            \addlegendentry{different, not sure}; \addlegendentry{different, absolutely sure}
            \nextgroupplot[
                    title={LibriTTS}, 
                    height=5cm, width=5cm,
                    ybar stacked,
                    ymin=0, ymax=100,
                    xtick=data,
                    symbolic x coords={AdaIN-VC, \textsc{AutoVC}, \textsc{AutoVC}-Vvec, Ground truth},
                    x tick label style={
                            rotate=45,
                            anchor=east,
                            font=\footnotesize
                        },
                    xtick style={
                            draw=none,
                        },
                ]
            \addplot[
                    ybar,
                    fill=blue,
                ] coordinates {
                    (AdaIN-VC, 16.4556962)
                    (\textsc{AutoVC}, 02.479338843)
                    (\textsc{AutoVC}-Vvec, 07.264957265)
                    (Ground truth, 41.35021097)
                };
            \addplot[
                    ybar,
                    fill=teal,
                ] coordinates {
                    (AdaIN-VC, 25.73839662)
                    (\textsc{AutoVC}, 08.26446281)
                    (\textsc{AutoVC}-Vvec, 20.51282051)
                    (Ground truth, 22.78481013)
                };
            \addplot[
                    ybar,
                    fill=orange,
                ] coordinates {
                    (AdaIN-VC, 22.78481013)
                    (\textsc{AutoVC}, 21.07438017)
                    (\textsc{AutoVC}-Vvec, 35.47008547)
                    (Ground truth, 22.3628692)
                };
            \addplot[
                    ybar,
                    fill=red,
                ] coordinates {
                    (AdaIN-VC, 35.02109705)
                    (\textsc{AutoVC}, 68.18181818)
                    (\textsc{AutoVC}-Vvec, 36.75213675)
                    (Ground truth, 13.5021097)
                };
        \end{groupplot}
    \end{tikzpicture}
    }
      {\pgfplotslegendfromname{fred}}
    \caption{Subjective evaluation results on speaker similarity of AdaIN-VC, \textsc{AutoVC} and \textsc{AutoVC}-Vvec.}
    \label{fig:subjective_similarity_results}
\end{figure}
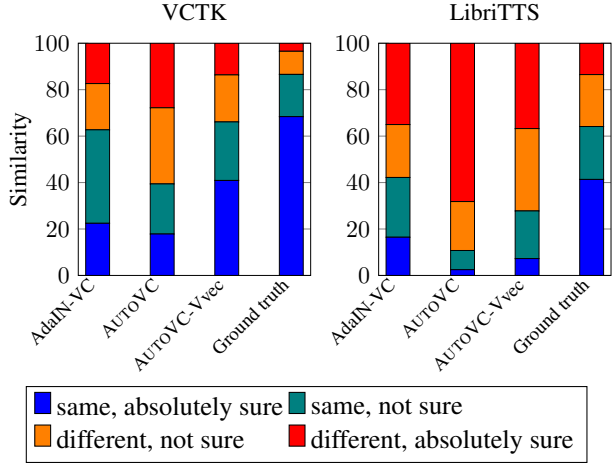

\begin{figure}[htb]
    \resizebox{.95\linewidth}{!}{
    \begin{tikzpicture}
        \begin{groupplot}[
                group style={
                        group size=2 by 1,
                    },
            ]
            
            \nextgroupplot[
                    title={VCTK}, 
                    height=5cm, width=5 cm,
                    ylabel={MOS Score},
                    ylabel style={
                            yshift=-1mm,
                        },
                    ymin=0, ymax=5.5,
                    xtick=data,
                    symbolic x coords={AdaIN-VC, AutoVC, AutoVC-Vvec, Ground truth},
                    x tick label style={
                            rotate=45,
                            anchor=east,
                            font=\footnotesize
                        },
                    xtick style={
                            draw=none,
                        },
                    nodes near coords,
                    nodes near coords align={vertical},
                    nodes near coords style={
                            inner sep=3pt,
                            /pgf/number format/.cd,
                            fixed,
                            fixed zerofill,
                            precision=2,
                        },
                ]
            \addplot[
                    ybar,
                    fill=orange,
                    error bars/.cd,
                    error bar style={line width=2pt},
                    y dir=both,
                    y explicit,
                ] coordinates {
                    (AdaIN-VC, 2.96) +- (0, 0.1405937785)
                    (AutoVC, 3.428)  +- (0, 0.1448155305) 
                    (AutoVC-Vvec, 3.392) +- (0, 0.1370482166)
                    (Ground truth, 4.516) +- (0, 0.09487067954)
                };
                
            \nextgroupplot[
                    title={LibriTTS}, 
                    height=5cm, width=5 cm,
                    ymin=0, ymax=5.5,
                    xtick=data,
                    symbolic x coords={AdaIN-VC, AutoVC, AutoVC-Vvec, Ground truth},
                    x tick label style={
                            rotate=45,
                            anchor=east,
                            font=\footnotesize
                        },
                    xtick style={
                            draw=none,
                        },
                    nodes near coords,
                    nodes near coords align={vertical},
                    nodes near coords style={
                            inner sep=3pt,
                            /pgf/number format/.cd,
                            fixed,
                            fixed zerofill,
                            precision=2,
                        }
                ]
            \addplot[
                    ybar,
                    fill=orange,
                    error bars/.cd,
                    error bar style={line width=2pt},
                    y dir=both,
                    y explicit,
                ] coordinates {
                    (AdaIN-VC, 2.752) +- (0, 0.142552642)
                    (AutoVC, 2.56) +- (0, 0.1482541427)
                    (AutoVC-Vvec, 2.912) +- (0, 0.1424318653)
                    (Ground truth, 4.568) +- (0, 0.09644743602)
                };
            
        \end{groupplot}
    \end{tikzpicture}
    }
    \caption{Subjective evaluation results on naturalness of AdaIN-VC, \textsc{AutoVC} and \textsc{AutoVC}-Vvec.}
    \label{fig:subjective_naturalness_results}
\end{figure}
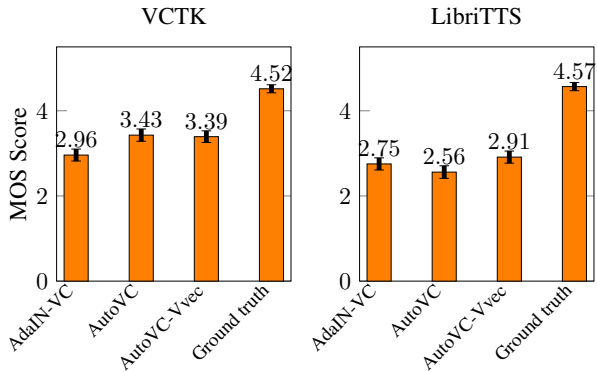

\subsection{Experimental setup}
We trained models with three different pretrained speaker embeddings, including i-vector, d-vector and x-vector denoted as \textbf{-I}, \textbf{-D} and \textbf{-X} respectively.
We denote VC models whose speaker encoder is jointly trained as \textbf{-Joint}.
In addition, we trained \textsc{AutoVC} whose speaker encoder is replaced with v-vector from trained AdaIN-VC (\textbf{\textsc{AutoVC}-Vvec}).

\subsection{Objective results}
The results are shown in Fig.~\ref{fig:The Influence of Speaker Embeddings}.
We can see AdaIN-VC-Joint performed the best among all AdaIN-VC modifications, so did \textsc{AutoVC}-Vvec among all \textsc{AutoVC} modifications.
The models trained with d-vector and x-vector were not capable of converting speaker timbre well.
The models trained with i-vector were able to convert timbre well but with unsatisfactory CER.
V-vector was trained on far less data than other pretrained speaker embeddings did (VCTK versus VoxCeleb), but still outperformed pretrained ones.
Finally, \textsc{AutoVC}-Joint failed, with extremely high CER and low SVAR in each case.

The results suggest that pretrained speaker embeddings are not as effective for VC as they are in speaker verification although they are pretrained with a large number of speakers.
Speaker embedding jointly trained with VC model is more suitable for voice conversion than those trained on speaker verification.
Moreover, the results also indicate that the speaker information required for VC and speaker verification are different.
Finally, the result also show that \textsc{AutoVC} needs a pretrained speaker embedding so as to make content encoder able to extract content information only.

\subsection{Subjective results}
We conducted Mean Opinion Score (MOS) test to rate the quality and similarity of the converted utterances.
For speech quality, subjects were asked to score the utterances based on its naturalness from 1 (very bad) to 5 (very good).
As for speaker similarity, subjects were given two utterances each time and needed to score based on their similarity from 1 (different, absolutely sure) to 4 (same, absolutely sure).
All of the MOS results are reported with 95\% confidence intervals.

We compare top three models in the previous experiments, including AdaIN-VC, \textsc{AutoVC} and \textsc{AutoVC} with pretrained speaker encoder trained by AdaIN-VC (\textsc{AutoVC}-Vvec).
Models were evaluated on two corpora, VCTK-train and LibriTTS.
Both source and target utterances were from the same corpus.
For each model, we selected 50 examples from VCTK-train and LibriTTS respectively, and each example was scored by at least 5 subjects.

The results of subjective evaluation are similar to those of objective evaluations, reinforcing our results in the last few sections.
Both from the MOS and similarity tests, we can see that \textsc{AutoVC}-Vvec mitigates the problem that \textsc{AutoVC} performs undesirably when faced with mismatched data distribution.
The results also suggest that utilizing v-vector improves on both objective and subjective metrics.

%% file: 7_Conclusion.tex
\section{Conclusion}
We performed a survey on the robustness of current voice conversion models in several perspectives.
The sampling rate and audio duration have great impact on these models.
In addition, carefully-designed information bottleneck in \textsc{AutoVC} does not generalize well to unseen data, whereas AdaIN-VC is relatively more robust.
Also, speaker embeddings jointly trained with other VC modules are more suitable for the conversion task than those pre-trained for speaker identification.
Last but not least, objective and subjective results are highly correlated, indicating that the objective metrics adopted here are appropriate measures for voice conversion.

%% file: 8_Acknowledgement.tex
\section{Acknowledgement}
We thank to National Center for High-performance Computing (NCHC) for providing computational and storage resources.

%% file: Template_Regular.bbl
\begin{thebibliography}{10}

\bibitem{1810.12656}
Li-Wei Chen, Hung-Yi Lee, and Yu~Tsao,
\newblock ``Generative adversarial networks for unpaired voice transformation
  on impaired speech,'' 2018.

\bibitem{8553236}
T.~{Kaneko} and H.~{Kameoka},
\newblock ``Cyclegan-vc: Non-parallel voice conversion using cycle-consistent
  adversarial networks,''
\newblock in {\em 2018 26th European Signal Processing Conference (EUSIPCO)},
  2018, pp. 2100--2104.

\bibitem{serra2019blow}
Joan Serr{\`a}, Santiago Pascual, and Carlos~Segura Perales,
\newblock ``Blow: a single-scale hyperconditioned flow for non-parallel
  raw-audio voice conversion,''
\newblock in {\em Advances in Neural Information Processing Systems}, 2019, pp.
  6793--6803.

\bibitem{2006.04154}
Da-Yi Wu, Yen-Hao Chen, and Hung-Yi Lee,
\newblock ``Vqvc+: One-shot voice conversion by vector quantization and u-net
  architecture,'' 2020.

\bibitem{7820786}
C.~{Hsu}, H.~{Hwang}, Y.~{Wu}, Y.~{Tsao}, and H.~{Wang},
\newblock ``Voice conversion from non-parallel corpora using variational
  auto-encoder,''
\newblock in {\em 2016 Asia-Pacific Signal and Information Processing
  Association Annual Summit and Conference (APSIPA)}, 2016, pp. 1--6.

\bibitem{chou2018multi}
Ju-chieh Chou, Cheng-chieh Yeh, Hung-yi Lee, and Lin-shan Lee,
\newblock ``Multi-target voice conversion without parallel data by
  adversarially learning disentangled audio representations,''
\newblock {\em Proc. Interspeech 2018}, pp. 501--505, 2018.

\bibitem{kameoka2018stargan}
Hirokazu Kameoka, Takuhiro Kaneko, Kou Tanaka, and Nobukatsu Hojo,
\newblock ``Stargan-vc: Non-parallel many-to-many voice conversion using star
  generative adversarial networks,''
\newblock in {\em 2018 IEEE Spoken Language Technology Workshop (SLT)}. IEEE,
  2018, pp. 266--273.

\bibitem{Liu2018}
Songxiang Liu, Jinghua Zhong, Lifa Sun, Xixin Wu, Xunying Liu, and Helen Meng,
\newblock ``Voice conversion across arbitrary speakers based on a single
  target-speaker utterance,''
\newblock in {\em Proc. Interspeech 2018}, 2018, pp. 496--500.

\bibitem{inproceedings_2}
Andy Liu, Po-chun Hsu, and Hung-yi Lee,
\newblock ``Unsupervised end-to-end learning of discrete linguistic units for
  voice conversion,''
\newblock 09 2019, pp. 1108--1112.

\bibitem{unknown}
Kaizhi Qian, Zeyu Jin, Mark Hasegawa-Johnson, and Gautham Mysore,
\newblock ``F0-consistent many-to-many non-parallel voice conversion via
  conditional autoencoder,'' 04 2020.

\bibitem{inproceedings}
Hui Lu, Zhiyong Wu, Dongyang Dai, Runnan Li, Shiyin Kang, Jia Jia, and Helen
  Meng,
\newblock ``One-shot voice conversion with global speaker embeddings,''
\newblock 09 2019, pp. 669--673.

\bibitem{2004.11284}
Kaizhi Qian, Yang Zhang, Shiyu Chang, David Cox, and Mark Hasegawa-Johnson,
\newblock ``Unsupervised speech decomposition via triple information
  bottleneck,'' 2020.

\bibitem{chou2019one}
Ju-chieh Chou and Hung-Yi Lee,
\newblock ``One-shot voice conversion by separating speaker and content
  representations with instance normalization,''
\newblock {\em Proc. Interspeech 2019}, pp. 664--668, 2019.

\bibitem{qian2019autovc}
Kaizhi Qian, Yang Zhang, Shiyu Chang, Xuesong Yang, and Mark Hasegawa-Johnson,
\newblock ``Autovc: Zero-shot voice style transfer with only autoencoder
  loss,''
\newblock in {\em International Conference on Machine Learning}, 2019, pp.
  5210--5219.

\bibitem{6854363}
E.~{Variani}, X.~{Lei}, E.~{McDermott}, I.~L. {Moreno}, and
  J.~{Gonzalez-Dominguez},
\newblock ``Deep neural networks for small footprint text-dependent speaker
  verification,''
\newblock in {\em 2014 IEEE International Conference on Acoustics, Speech and
  Signal Processing (ICASSP)}, 2014, pp. 4052--4056.

\bibitem{veaux2016superseded}
Christophe Veaux, Junichi Yamagishi, Kirsten MacDonald, et~al.,
\newblock ``Superseded-cstr vctk corpus: English multi-speaker corpus for cstr
  voice cloning toolkit,''
\newblock 2016.

\bibitem{2005.08781}
Chien yu~Huang, Yist~Y. Lin, Hung yi~Lee, and Lin shan Lee,
\newblock ``Defending your voice: Adversarial attack on voice conversion,''
  2020.

\bibitem{1420369}
P.~{Kenny}, G.~{Boulianne}, and P.~{Dumouchel},
\newblock ``Eigenvoice modeling with sparse training data,''
\newblock {\em IEEE Transactions on Speech and Audio Processing}, vol. 13, no.
  3, pp. 345--354, 2005.

\bibitem{8461375}
D.~{Snyder}, D.~{Garcia-Romero}, G.~{Sell}, D.~{Povey}, and S.~{Khudanpur},
\newblock ``X-vectors: Robust dnn embeddings for speaker recognition,''
\newblock in {\em 2018 IEEE International Conference on Acoustics, Speech and
  Signal Processing (ICASSP)}, 2018, pp. 5329--5333.

\bibitem{goodfellow2014generative}
Ian Goodfellow, Jean Pouget-Abadie, Mehdi Mirza, Bing Xu, David Warde-Farley,
  Sherjil Ozair, Aaron Courville, and Yoshua Bengio,
\newblock ``Generative adversarial nets,''
\newblock in {\em Advances in neural information processing systems}, 2014, pp.
  2672--2680.

\bibitem{kingma2013auto}
Diederik~P Kingma and Max Welling,
\newblock ``Auto-encoding variational bayes,''
\newblock {\em arXiv preprint arXiv:1312.6114}, 2013.

\bibitem{huang2017arbitrary}
Xun Huang and Serge Belongie,
\newblock ``Arbitrary style transfer in real-time with adaptive instance
  normalization,''
\newblock in {\em Proceedings of the IEEE International Conference on Computer
  Vision}, 2017, pp. 1501--1510.

\bibitem{wan2018generalized}
Li~Wan, Quan Wang, Alan Papir, and Ignacio~Lopez Moreno,
\newblock ``Generalized end-to-end loss for speaker verification,''
\newblock in {\em 2018 IEEE International Conference on Acoustics, Speech and
  Signal Processing (ICASSP)}. IEEE, 2018, pp. 4879--4883.

\bibitem{zen2019libritts}
Heiga Zen, Viet Dang, Rob Clark, Yu~Zhang, Ron~J Weiss, Ye~Jia, Zhifeng Chen,
  and Yonghui Wu,
\newblock ``Libritts: A corpus derived from librispeech for text-to-speech,''
\newblock {\em Proc. Interspeech 2019}, pp. 1526--1530, 2019.

\bibitem{panayotov2015librispeech}
Vassil Panayotov, Guoguo Chen, Daniel Povey, and Sanjeev Khudanpur,
\newblock ``Librispeech: an asr corpus based on public domain audio books,''
\newblock in {\em 2015 IEEE International Conference on Acoustics, Speech and
  Signal Processing (ICASSP)}. IEEE, 2015, pp. 5206--5210.

\bibitem{kominek2004cmu}
John Kominek and Alan~W Black,
\newblock ``The cmu arctic speech databases,''
\newblock in {\em Fifth ISCA workshop on speech synthesis}, 2004.

\bibitem{wang2015thchs}
Dong Wang and Xuewei Zhang,
\newblock ``Thchs-30: A free chinese speech corpus,''
\newblock {\em arXiv preprint arXiv:1512.01882}, 2015.

\bibitem{griffin1984signal}
Daniel Griffin and Jae Lim,
\newblock ``Signal estimation from modified short-time fourier transform,''
\newblock {\em IEEE Transactions on Acoustics, Speech, and Signal Processing},
  vol. 32, no. 2, pp. 236--243, 1984.

\bibitem{van_den_Oord+2016}
Aäron van~den Oord, Sander Dieleman, Heiga Zen, Karen Simonyan, Oriol Vinyals,
  Alex Graves, Nal Kalchbrenner, Andrew Senior, and Koray Kavukcuoglu,
\newblock ``Wavenet: A generative model for raw audio,''
\newblock in {\em 9th ISCA Speech Synthesis Workshop}, 2016, pp. 125--125.

\bibitem{kumar2019melgan}
Kundan Kumar, Rithesh Kumar, Thibault de~Boissiere, Lucas Gestin, Wei~Zhen
  Teoh, Jose Sotelo, Alexandre de~Br{\'e}bisson, Yoshua Bengio, and Aaron~C
  Courville,
\newblock ``Melgan: Generative adversarial networks for conditional waveform
  synthesis,''
\newblock in {\em Advances in Neural Information Processing Systems}, 2019, pp.
  14910--14921.

\bibitem{Povey:192584}
Daniel Povey, Arnab Ghoshal, Gilles Boulianne, Lukas Burget, Ondrej Glembek,
  Nagendra Goel, Mirko Hannemann, Petr Motlicek, Yanmin Qian, Petr Schwarz, Jan
  Silovsky, Georg Stemmer, and Karel Vesely,
\newblock ``The kaldi speech recognition toolkit,''
\newblock 2011,
\newblock IEEE Catalog No.: CFP11SRW-USB.

\bibitem{Nagrani17}
A.~Nagrani, J.~S. Chung, and A.~Zisserman,
\newblock ``Voxceleb: a large-scale speaker identification dataset,''
\newblock in {\em INTERSPEECH}, 2017.

\bibitem{Chung18b}
J.~S. Chung, A.~Nagrani, and A.~Zisserman,
\newblock ``Voxceleb2: Deep speaker recognition,''
\newblock in {\em INTERSPEECH}, 2018.

\bibitem{doi:10.1111/j.1749-6632.1987.tb48734.x}
MICHAEL~F. SHLESINGER,
\newblock ``Fractal time and 1/f noise in complex systems,''
\newblock {\em Annals of the New York Academy of Sciences}, vol. 504, no. 1,
  pp. 214--228, 1987.

\end{thebibliography}
